\documentclass[twocolumn]{aa}

\usepackage{graphicx,txfonts,natbib,verbatim}
\bibpunct{(}{)}{;}{a}{}{,}

\begin{document}

\title{Parallel electric field generation by Alfven wave turbulence}

\author{N.H. Bian, E. P. Kontar  \and J. C. Brown}

\offprints{N.H. Bian \email{nbian@astro.gla.ac.uk}}

\institute{Department of Physics \& Astronomy, University of Glasgow, G12 8QQ, United Kingdom}

\date{Received ; Accepted }

\abstract{}{This work aims to investigate the spectral structure of the parallel electric field
generated by strong anisotropic and balanced Alfvenic turbulence in relation
with the problem of electron acceleration from the thermal population in solar flare plasma conditions.}
{We consider anisotropic Alfvenic fluctuations in
the presence of a strong background magnetic field. Exploiting this anisotropy, a set of reduced equations governing
non-linear, two-fluid plasma dynamics is derived. The low-$\beta$ limit of this model is used to follow the turbulent cascade
of the energy resulting from the non-linear interaction between kinetic Alfven waves, from the large
magnetohydrodynamics (MHD) scales with $k_{\perp}\rho_{s}\ll 1$ down to the small "kinetic" scales with $k_{\perp}\rho_{s} \gg 1$, $\rho_{s}$ being
the ion sound gyroradius.}
{Scaling relations are obtained for the magnitude of the turbulent electromagnetic fluctuations, as a function of $k_{\perp}$ and $k_{\parallel}$,
showing that the electric field develops
a component parallel to the magnetic field at large MHD scales.}
{The spectrum we derive for the parallel electric field fluctuations
can be effectively used to model stochastic resonant acceleration and heating of electrons by Alfven waves
in solar flare plasma conditions}


\titlerunning{Parallel electric field generation by kinetic Alfven wave turbulence}

\authorrunning{Bian et al}

\maketitle

\section{Introduction}

Solar flares provide many challenges
for crucial aspects of high energy astrophysics, including
energy release, particle acceleration and transport
in magnetized plasmas \citep[e.g.][as recent reviews]{Aschwanden2002,Brown_etal2006}.
The impulsive phase of a flare marks the rapid
release and conversion of a large amount of magnetic energy, stored
in the solar corona, into the kinetic energy of particles.
In the standard thick-target model, \citep{Brown1971,Syrovatskii1972,LinHudson1976},
reviewed by \citep{Brown_etal2003,BrownKontar2005}, the stream of fast electrons which
emits bremsstrahlung hard X-rays heats the dense chromospheric plasma collisionally,
is produced first in the tenuous corona by electron acceleration from thermal
energies ($\lesssim 1$ keV) to deka-keV and MeV energies.
This standard geometry of flare electron acceleration and transport is consistent with a variety of spatially resolved
observations \citep[]{Aschwanden_etal2002,Emslie_etal2003,Kontar_etal2008,KruckerLin2008} and by electron time of flight effects in Hard X-ray light curves \citep{Aschwanden2002}. However, an electron beam undergoing solely collisional energy loss, as in the standard thick target model, gives
up around $10^5$ times energy to heat than to bremsstrahlung and demands \citep{Brown1971} a very high electron production rate to yield observed hard X-ray fluxes.
Furthermore the electron beam and hard X-ray source anisotropies
in the standard thick target model \citep{Brown1972} are much higher than inferred
from the flare hard X-ray data \citep{KontarBrown2006}. \citet{Brown_etal2009} have proposed
that if fast electrons,  on reaching the chromosphere, undergo re-acceleration by current sheets
there, their enhanced lifetimes increase the hard X-ray yield per electron, so reducing
the injection rate needed for hard X-ray production, while greatly reducing the fast
electron anisotropy in the main hard X-ray source. Therefore, any mechanism that can
re-accelerate electrons in the chromosphere is also of interest.

Various acceleration mechanisms have been proposed for energetic solar particles \citep{Aschwanden2002}, including
acceleration by a large scale parallel electric field \citep{Holman1985},
electric fields inside current sheets \citep{Litv2003,WoodNeukirch2005,Bian2008,SiverskyZharkova2009},
collapsing trap acceleration \citep{BogachevSomov2007}
as well as turbulent non-resonant \citep[]{BykovFleishman2009},
and resonant acceleration by waves \citep[see the reviews by][]{Miller_etal1997,Petrosian1999}.
Parallel acceleration by resonant interaction between electrons and the parallel electric
field produced by turbulent Alfven waves is the subject of the present study.

The resonant coupling between a given electromagnetic mode
characterized by its dispersion relation $\omega (\mathbf{k})$ and an electron gyrating
at the gyrofrequency $\omega_{ce}=qB_{0}/m$ while streaming at the speed $v_{\parallel}$
along the magnetic field, is given by the Doppler resonance condition,
$\omega-s\omega_{ce}/\gamma=k_{\parallel}v_{\parallel}$.
In this expression,
$k_{\parallel}$ is the parallel wavenumber of the wave,
$\gamma$ is the Lorentz factor and $s$ is the harmonic number of $\omega_{ce}$. Basically, the resonance condition specifies under which condition
this electron experiences an electromagnetic force which is
stationary. Therefore, if a broad spectrum of the electromagnetic field fluctuations associated with a particular mode is present,
and moreover, if the resonance condition with this mode is satisfied for thermal electrons, then
it is possible for these electrons to achieve a large energy gain, only limited by the
final energy which corresponds to the last resonance with this mode.
Within quasilinear theory, this resonant acceleration process is a diffusion in velocity space, from the
thermal velocity $V_{Te}$ up to the final velocity $V_{f}$.
The most straightforward way of producing a stream of fast electrons
accelerated along the magnetic field lines is through wave resonance satisfying the condition
\begin{equation}
\omega=k_{\parallel}v_{\parallel},
\label{eq:landau_res}
\end{equation}
either by the parallel electric force $F=qE_{\parallel}$ or the magnetic mirror force
$F=\mu \nabla_{\parallel} B$, $\mu= m v^{2}_{\perp}/2B_{0}$ being the magnetic moment.

Many people starting from \citep{Fermi1949} considered stochastic
acceleration of particles. \citet{Miller_etal1996} have developed a model
of thermal electron acceleration during flares based on the Landau resonance
between these electrons and the fluctuating parallel mirror force produced by
the compressive magnetic field component of turbulent magnetoacoustic waves.
The mechanism being the magnetic analog of Landau damping is called transit-time damping.
Since magnetoacoustic waves have similar speeds as Alfven waves,
their frequency being given by $\omega=kV_{a}$
they indeed can resonate with a population
of thermal electrons, i.e. $V_{Te}\sim V_A$. Under typical plasma conditions in the solar
corona \citep[e.g.][]{Emslie_etal2003,Kontar_etal2008},
i.e. magnetic field $B_{0}\approx 100$~G, plasma
density $n\approx 5\times 10^{9}cm^{-3}$, and electron
temperature $T_e\approx 10^{6}$~K, the Alfven velocity
($V_{A}\sim 3\times 10^8$~cm/s)
is close to the electron thermal speed ($V_{Te}\sim 4\times 10^{8}$~cm/s).
In the model by \citet{Miller_etal1996}, the broad spectrum of magnetic fluctuations
is produced by isotropic MHD turbulence.

As stated above, the Landau resonance (\ref{eq:landau_res}) is well satisfied between
thermal electrons and shear-Alfven waves with frequency given by $\omega=k_{\parallel}V_{A}$.
However, it is often assumed in the literature that
the shear-Alfven mode lacks the parallel electric field necessary to accelerate the particles.
This is only true if non-MHD effects are ignored in the range of wavenumbers where the wave
has a frequency $\omega\sim k_{\parallel}V_{A}$.
In this study, we reconsider the possibility of electron acceleration through the
Landau resonance with the fluctuating parallel electric force produced by Alfvenic turbulence. This is done
by investigating the spectral structure of the parallel electric field fluctuation resulting from
kinetic Alfven wave (KAW) turbulence, the KAW mode having a frequency given by $\omega=k_{\parallel}V_{A}(1+\rho_{s}^{2}k_{\perp}^{2})^{1/2}$ where
$\rho_{s}$ is the ion-sound gyro radius. Following the same lines as the Goldreich-Sidhrar theory for Alfvenic turbulence \citep{GoldreichSridhar1995},
we derive an expression for the parallel electric field spectrum,
for strong anisotropic KAW
turbulence, from the large MHD scales with $k_{\perp}\rho_{s}\ll 1$ down to the small "kinetic" scales with $k_{\perp}\rho_{s} \gg 1$. It is shown that the magnitude of the
the parallel electric field fluctuation, being an increasing function
of wave number in the MHD regime but a decreasing function of the wave number in the "kinetic" regime, it reaches a maximum
at the boundary, where the Alfven wave becomes dispersive. This means that the condition $k_{\perp}\rho_{s} \gg 1$ does not have to be satisfied
for stochastic acceleration by Alfven waves to be effective.

\section{Two-fluid plasma dynamics}

The starting point is a reduced set of equations,
describing anisotropic two-fluid plasma dynamics in a
strong magnetic field. Under the assumption of quasi-neutrality and
considering that the bulk plasma electrons have negligibly
small inertia, the fluid equations of motion
for the ions and the electrons are
\begin{equation}
nm_{i}(\partial_{t}\mathbf{V}_{i}+\mathbf{V}_{i}.\nabla\mathbf{V}_{i})=-\nabla
P_{i}+ne(\mathbf{E}+\mathbf{V}_{i}\times\mathbf{B}),
\label{mhd_i}
\end{equation}
\begin{equation}
0=-\nabla
P_{e}-ne(\mathbf{E}+\mathbf{V}_{e}\times\mathbf{B}),
\label{mhd_e}
\end{equation}
where $n$ is the plasma number density, $\mathbf{V}_{i/e}$ is the ion/electron velocity,
$m_{i}$ is the ion mass, $P_{i,e}$ the ion/electron pressure, $\mathbf{E}$ is the electric
field and $\mathbf{B}$ the magnetic field.
The system should be supplemented by Maxwell's equations :
$\nabla\times \mathbf{E}=-\partial_{t}\mathbf{B}$, $\nabla\times \mathbf{B}=\mu_{0}\mathbf{j}$
and $\nabla.\mathbf{B}=0$ with $\mathbf{j}=ne(\mathbf{V}_{i}-\mathbf{V}_{e})$.

These equations are made dimensionless by introducing a typical length scale
$L_{0}$, density $n_{0}$, a typical value for the magnetic field $B_{0}$,
corresponding to the Alfven velocity $V_{A}=B_{0}/\sqrt{(\mu_{0}nm_{i})}$,
a time scale $L_{0}/V_{A}$ and the pressures are normalized to the magnetic
pressure $B_{0}^{2}/\mu_{0}$. Equations (\ref{mhd_i}-\ref{mhd_i}) are then combined
to give an ion equation of motion
\begin{equation}\label{mot}
\partial_{t}\mathbf{V}+\mathbf{V}.\nabla\mathbf{V}=-\nabla P+\mathbf{j}\times\mathbf{B},
\end{equation}
with $P\equiv P_{i}+P_{e}$, while the electron equation of
motion is equivalent to the generalized Ohm's law,
\begin{equation}\label{ol}
\mathbf{E}+\mathbf{V}_{e}\times\mathbf{B}=-d_{i}\nabla P_{e},
\end{equation}
with $\mathbf{V}_{e}=\mathbf{V}-d_{i}\mathbf{j}$ and $\mathbf{V}\equiv \mathbf{V}_{i}$.

Ohm's law (\ref{ol}) involves the non-dimensional
parameter $d_{i}$ which is the normalized ion skin depth
$d_{i}\equiv (c/\omega_{pi})/L_{0}$ with $\omega_{pi}=\sqrt{(ne^{2}/\epsilon_{0}m_{i})}$.
The expression for its magnetic field aligned component,
\begin{equation}
E_{\parallel}=-d_{i}\nabla_{\parallel}P_{e},
\end{equation}
shows that a parallel electric field can be produced by the electron
pressure gradient along the magnetic field lines. We emphasize that this electric field is parallel to the total
magnetic field, comprising the background plus its perturbation.

The existence of a strong background magnetic field $B_{0}\mathbf{z}$
makes the plasma dynamics anisotropic with $\alpha \equiv k_{\parallel}/k_{\perp}\ll 1$.
We can write the normalized magnetic field as
$\mathbf{B}=\mathbf{z}+\delta \mathbf{B}$ and make
the following ordering $\delta B\sim \alpha$ for its perturbation. The solenoidal condition for
the magnetic field perturbation allows its perpendicular
component to be written in term of a flux
function: $\delta\mathbf{B}\simeq\nabla \psi\times \mathbf{z}+b_{z}\mathbf{z}$.
In the same way, the perpendicular velocity is written
in term of a stream function: $\mathbf{V}\simeq\nabla \phi\times \mathbf{z}+v_{z}\mathbf{z}$ with the ordering
$V\sim \delta B$.
Following the same standard procedure as is employed to obtain reduced
magnetohydrodynamics (RMHD) from the MHD
equations \citep{KadomtsevPogutse1974,Strauss1976}, the two-fluid equations (\ref{mot})-(\ref{ol}) yield, to
order $\alpha^{2}$ in the asymptotic expansion,
\begin{equation}\label{eq16}
\partial_{t}\psi=\partial_{z}(\phi-d_{i}b_{z})+ [\phi-d_{i}b_{z}, \psi],
\end{equation}
\begin{equation}\label{eq17}
\partial_{t} b_{z}=\partial_{z}(v_{z}-d_{i}j_{z})+[v_{z}-d_{i}j_{z},\psi]+[\phi,b_{z}]-\nabla.\mathbf{V},
\end{equation}
\begin{equation}\label{eq18}
\partial_{t} \omega_{z}=\partial_{z} j_{z}+[j_{z},\psi]+[\phi,\omega_{z}],
\end{equation}
\begin{equation}\label{eq19}
\partial_{t} v_{z}=\partial _{z}b_{z}+[b_{z},\psi]+[\phi, v_{z}],
\end{equation}
where the notation $[A,B]=\mathbf{z}.(\nabla A \times \nabla B)$ is
adopted and $j_{z}=-\nabla^{2}\psi$, $\omega_{z}=-\nabla^{2}\phi$
are respectively the $z$-component of the current and the vorticity.
In the reduction scheme, the fast time scale is the propagation time
of the fast magnetoacoustic mode, which is therefore eliminated,
while the low-frequency dynamics of the shear Alfv\'en and slow magnetoacoustic
modes are retained. The above system can be closed by the pressure equation,
\begin{equation}\label{eq20}
\partial_{t} p=[\phi,p]-\beta\nabla.\mathbf{V},
\end{equation}
with $p$ the normalized pressure perturbation.
The plasma pressure parameter is defined
as $\beta= C_{s}^{2}/V_{A}^{2} =\Gamma P_{0}/(B_{0}^{2}/\mu_{0})$,
with $\Gamma$ the ratio of specific heats and $P_{0}$ the background reference pressure.

In the limit $\beta\gg 1$, it can be seen from (\ref{eq20}) that the plasma
flow becomes incompressible, $\nabla.\mathbf{V}=0$, and hence,
the above system is equivalent to the incompressible reduced Hall-MHD equations derived,
for instance, by \citet{Gomez_etal2008}.
For $d_{i}=0$, the classical reduced-MHD equations are recovered with
equations (\ref{eq16}) and (\ref{eq18}) forming an independent system
describing the non-linear dynamics of shear Alfv\'en waves.

Assuming $\mathbf{V}=0$ in the previous model leads to the reduced
electron-MHD (EMHD) equations which conserve the magnetic
energy $E=\int d^{3}r[(\nabla \psi)^{2}+b_{z}^{2}].$ Its linear modes are the whistler
waves with $\omega_{\pm}=\pm k_{\parallel}d_{i}k_{\perp}$.
Some key properties of the EMHD turbulence have been investigated
both numerically and theoretically, suggesting that the Kolmogorov type arguments
work fine \citep{Biskamp_etal1999,Ng_etal2003, ChoLazarian2004,Cho2009}. A calculation along the lines
of the one below for KAWs, shows that the energy spectrum for whistler wave
turbulence is $E_{k_{\perp}}\propto \epsilon^{2/3}d_{i}^{-2/3}k_{\perp}^{-7/3}$.

Relaxing the assumption of a large $\beta$, we can allow for the effect
of a finite plasma compressibility. Since the perpendicular pressure balance,
$\nabla_{\perp}(p+b_{z})=0 $, is satisfied to order
$\alpha$ in the expansion of the ion equation of motion, the compression
term $\nabla.\mathbf{V}$ can be eliminated from (\ref{eq20}) and (\ref{eq17}) using
the fact that $p\simeq b_{z}$. Therefore, defining $Z=b_{z}/c_{\beta}$,
$c_{\beta}=\sqrt{\beta/1+\beta}$ and $d_{\beta}=c_{\beta}d_{i}$, the following model
is obtained :
\begin{equation}\label{eq21b}
\partial_{t}\psi=\partial_{z}(\phi-d_{\beta}Z)+ [\phi-d_{\beta}Z,
\psi]
\end{equation}
\begin{equation}\label{eq22b}
\partial_{t} Z=\partial_{z}(c_{\beta}v_{z}-d_{\beta}j_{z})+[c_{\beta}v_{z}-d_{\beta}j_{z},\psi]+[\phi,Z]
\end{equation}
\begin{equation}\label{eq23b}
\partial_{t} \omega_{z}=\partial_{z} j_{z}+[j_{z},\psi]+[\phi,\omega_{z}]
\end{equation}
\begin{equation}\label{eq24b}
\partial_{t} v_{z}=c_{\beta}\partial _{z}Z+[\phi, v_{z}]+c_{\beta}[Z,\psi]
\end{equation}
The system conserves the total energy $E=\int d^{3}r[(\nabla\phi)^{2}+v^{2}_{z}+(\nabla \psi)^{2}+Z^{2}].$
More details concerning the derivation of this reduced two-fluid MHD model
can be found in \citep{BianTsiklauri2009}.

\section{Kinetic Alfven turbulence}
For $\beta\ll 1$, the parallel flow dynamics decouples in the above reduced two-fluid MHD model, hence
$Z=-\rho_{s}\omega_{z}$, and therefore it simplifies to:
 \begin{equation}\label{eq32}
\partial_{t}\psi=\partial_{z}(\phi+\rho^{2}_{s}\omega_{z})+ [\phi+\rho^{2}_{s}\omega_{z},
\psi],
\end{equation}
\begin{equation}\label{eq33}
\partial_{t} \omega_{z}=\partial_{z} j_{z}+[\phi,\omega_{z}]+[j_{z},\psi],
\end{equation}
with $\rho_{s}=C_{s}/\omega_{ci}$ being the ion sound gyroradius.
The total energy takes the form:
\begin{equation}\label{en}
E=\int d^{3}r[(\nabla\phi)^{2}+(\nabla \psi)^{2}+\rho_{s}^{2}\omega^{2}_{z}].
\end{equation}
Notice that this model is very similar to the EMHD system when $k_{\perp}\rho_{s}\gg 1$
, with the compressibility effect retained, while it reduces to the standard RMHD description of shear Alfv\'en waves perturbations
for $k_{\perp}\rho_{s}\ll 1$. Linearizing this two-field model yields the frequency
of the kinetic Alfven wave:
\begin{equation}\label{dr}
\omega_{\pm}=\pm k_{\parallel}\sqrt{1+\rho_{s}^{2}k^{2}_{\perp}}.
\end{equation}
This shows that the low-frequency Alfven wave, with a frequency much smaller that the
ion cyclotron frequency $\omega<\omega_{ci}$, becomes
dispersive when the wavelength perpendicular to the background magnetic field
is comparable or smaller than the ion sound gyroradius $\rho_{s}$, i.e.
$\omega_{\pm}=\pm k_{\parallel}\rho_{s}k_{\perp}$, this dispersion being similar to
the one of the low-frequency whistler wave.
We further notice that the model given by (\ref{eq32})-(\ref{eq33}) is the simplest subset
of the so-called electromagnetic gyrofluid models (see e.g. \citet{Waelbroeck_etal2009} and references therein) which are obtained as moments
of the drift-kinetic equations.

From Equations (\ref{eq32})-(\ref{eq33}), a theory for KAW turbulence is now constructed
along the same lines as the Goldreich-Sidrar theory \citep{GoldreichSridhar1995}
for Alfven wave turbulence \citep{Kraichnan1965}. Some form of dissipation at small
scales, balancing the energy input at large scales, is necessary for a steady cascade of energy to take place.
It is assumed that the turbulent
fluctuations are composed of KAWs, hence,
\begin{equation}\label{k}
\phi=\psi \left(\sqrt{\frac{1}{1+\rho_{s}^{2}k_{\perp}^{2}}}\right).
\end{equation}
Focusing first on the perpendicular cascade, we can express the energy per wave number
$E_{k_{\perp}}$ from (\ref{en}) and use (\ref{k}) to obtain that
\begin{equation}\label{en2}
E_{k_{\perp}}\propto k_{\perp}\psi^{2}_{k_{\perp}}.
\end{equation}
Moreover, we adopt the standard assumption that the flux of turbulent energy
at a given scale is determined by the turbulence at that scale and is a constant
equal to the energy injection rate $\epsilon$. Therefore, the expression for the energy cascade rate is
\begin{equation}\label{rat2}
\epsilon \sim k_{\perp}E_{k_{\perp}}/\tau_{NL},
\end{equation}
with the non-linear time scale being given by
$\tau_{NL} \sim 1/k^{2}_{\perp}\phi_{k_{\perp}}(1+\rho_{s}^{2}k_{\perp}^{2})$,
which, using (\ref{k}), is equivalently expressed as
\begin{equation}\label{nl2}
\tau_{NL} \sim \frac{1}{k_{\perp}^{2}\psi_{k_{\perp}}\sqrt{1+\rho_{s}^{2}k_{\perp}^{2}}}
\end{equation}
Combining relations (\ref{en2})-(\ref{rat2})-(\ref{nl2}) yields the scaling law for the energy spectrum:
\begin{equation}\label{spec}
E_{k_{\perp}}=C\epsilon^{2/3}k_{\perp}^{-5/3}(1+\rho_{s}^{2}k^{2}_{\perp})^{-1/3},
\end{equation}
where $C$ is a constant of the order of unity \citep{Kraichnan1965}.
This expression recovers the spectrum of Alfvenic turbulence, in the limit $k_{\perp}\rho_{s}\ll1$,
i.e., $E_{k_{\perp}}\sim C\epsilon^{2/3}k_{\perp}^{-5/3}$,
while in the dispersive range, for $k_{\perp}\rho_{s}\gg1$, then, $E_{k_{\perp}}\sim C\epsilon^{2/3}\rho_{s}^{-2/3}k_{\perp}^{-7/3}$.
Implicit in the derivation of (\ref{spec}) is the assumption that the fraction of the energy
flux of Alfvenic turbulence which is transferred from the MHD scales
onto the dispersive scales is of the order unity. Notice that since from (\ref{en2}), the magnetic energy spectrum
is $E_{k_{\perp}}=k^{-1}_{\perp}\delta B^{2}_{\perp}$, then
(\ref{spec}) is also equivalent to the scaling relation
\begin{equation}\label{db}
\delta B_{\perp}=C^{1/2}\epsilon^{1/3}k_{\perp}^{-1/3}(1+\rho_{s}^{2}k_{\perp}^{2})^{-1/6}
\end{equation}
Now, we recall a fundamental ordering
used in the derivation of the two-fluid reduced MHD system (\ref{eq32})-(\ref{eq33}) :
\begin{equation}\label{ord}
\delta B_{\perp}\sim k_{\parallel}/k_{\perp},
\end{equation}
This ordering is not restrictive in the sense that we are interested in the inertial range and not in
 the outer scale of the Alfvenic turbulence, where $\delta B_{\perp}$ can be of order unity.
Using (\ref{db}), (\ref{ord}) provides the scale dependent anisotropy of the turbulence :
\begin{equation}\label{cb2}
k_{\parallel}(k_{\perp})\sim \epsilon^{1/3}k_{\perp}^{2/3}(1+\rho_{s}^{2}k_{\perp}^{2})^{-1/6},
\end{equation}
which recovers the original Goldreich-Sridhar critical balance relation $k_{\parallel}\propto k_{\perp}^{2/3}$,
for Alfven wave turbulence when $k_{\perp}\rho_{s}\ll1$, while in the dispersive
range, $k_{\parallel}\propto k_{\perp}^{1/3}$.

In fact, it would have been equivalent to argue, following \citet{GoldreichSridhar1995}, that the anisotropy
of the turbulence is fixed by the so-called critical balance condition,
i.e. to assume that the characteristic non-linear decorrelation time
is of the order of the inverse KAW frequency, i.e.
$\omega ^{-1}_{KAW}\sim \tau _{NL}$,
with $\omega$ given by Eq.(\ref{dr}). The scaling relations
obtained for the energy spectrum and anisotropy in the dispersive scales
of kinetic Alfven wave turbulence \citep{CranmerVanBallegooijen2003,Howes_etal2008,Schekochihin_etal2009}
are similar to the ones of EMHD turbulence \citep{Biskamp_etal1999,Ng_etal2003,ChoLazarian2004,Cho2009}.
Notice however that for an EMHD Ohm's law given by $\mathbf{E}=d_{i}\mathbf{j}\times \mathbf{B}$, whistlers
do not have a parallel electric field.

Before concluding this section, few comments are due. In deriving the energy spectrum for kinetic Alfven wave turbulence
we are relying on the existing theory developed by \citep{GoldreichSridhar1995} for strong anisotropic
and balanced Alfven turbulence. The same approach was followed by \citep{Schekochihin_etal2009} based on
a compressible EMHD model to describe the dispersive range of Alfven turbulence, see also \citep{CranmerVanBallegooijen2003}.
It is our framework to investigate the spectral structure of the turbulent parallel electric field.
This should however not suggest that there is one universal cascade of Alfvenic fluctuations.
Indeed, the previous arguments are based on the assumption that the turbulence is non cross-helical. In the MHD range, the effect of cross-helicity
on the cascade of the two Elsasser energies, i.e. imbalanced turbulence, was investigated by \citep{Lithwick2007, Beresnyak2008, Chandran2008,
perez2009}. Imbalanced turbulence is a more general situation but it is not yet clear how the imbalance affects the dispersive range.
Moreover it should be mentioned that MHD and EMHD turbulence can also be dominated by weak fluctuations, see \citep{Galtier2002, Galtier2003},
 weak Alfven turbulence producing different spectra and anisotropy than the case studied here. Also, it has been argued that "dynamic alignment"
 of velocity and magnetic fields result in spectra that are flatter than Kolmogorov\citep{boldyrev2006,mason2006,beres2006}.
 With these restrictions in mind we can now discuss the parallel electric field spectrum
 of Alfven turbulence, a potentially important issue, which to the best of our knowledge
 has not been investigated so-far.

\section{Parallel electric field spectrum}
As stated above, the dispersive nature of the dynamics of KAWs involves the production
of an electric field perturbation $\delta \mathbf{E}$ which possesses a component parallel
to the magnetic field. The two components of $\delta \mathbf{E}$ are related
to the magnitude of the perpendicular magnetic field perturbation $ \delta B_{\perp}$
through the relations:
\begin{equation}
\delta E_{\perp}=(1+k_{\perp}^{2}\rho_{s}^{2})^{-1/2}\delta B_{\perp},
\end{equation}
for the perpendicular component and
\begin{equation}
\delta E_{\parallel}=k_{\parallel}k_{\perp}\rho_{s}^{2}(1+k_{\perp}^{2}\rho_{s}^{2})^{-1/2}\delta B_{\perp},
\end{equation}
for the parallel component.
Using the scaling (\ref{db}) for $\delta B_{\perp}$, it follows that
\begin{equation}
\delta E_{\perp}=C^{1/2}\epsilon^{1/3}k_{\perp}^{-1/3}(1+\rho_{s}^{2}k^{2}_{\perp})^{-2/3}
\end{equation}
and
\begin{equation}
\delta E_{\parallel}=C^{1/2}\epsilon^{1/3}\rho_{s}^{2}k_{\parallel}k_{\perp}^{2/3}(1+\rho_{s}^{2}k^{2}_{\perp})^{-2/3}
\end{equation}
These equations, together with Eq.($\ref{cb2}$), determine the magnitude of the electric
field fluctuations produced by strong anisotropic and balanced kinetic Alfven wave turbulence as a function of wavenumber $k_{\parallel}$
or $k_{\perp}$. Of particular interest is the scaling of the parallel electric field as a function of the
parallel wave-number. It is easily shown that the parallel electric field behaves like
$\delta E_{\parallel}(k_{\parallel})\propto \rho_{s}^{2}k_{\parallel}^{2}$
in the MHD scales when $k_{\parallel}\ll \epsilon^{1/3}\rho_{s}^{-2/3}$ and like
$\delta E_{\parallel}(k_{\parallel})\propto \epsilon k_{\parallel}^{-1}$,
in the dispersive scale when $k_{\parallel}\gg \epsilon^{1/3}\rho_{s}^{-2/3}$. The magnitude of the parallel
electric field, in V$m^{-1}$, is plotted in Fig.1 versus $k_{\parallel}\rho_{s}$, for different
normalized energy injection rates $\epsilon=(\delta B_{\perp}/B_{0})^{2}$, $\delta B_{\perp}$ being here the
dimensional value of the magnetic perturbation at the injection scale (see below).

\section{Discussion and conclusions }

There exists an extensive literature on stochastic acceleration by resonant interaction
between waves and particles. However, to our knowledge, a scenario
for field-aligned acceleration of electrons by the parallel electric field
produced by Alfvenic turbulence has not been
considered so-far. The reason is that it is generally assumed
that the MHD Alfven mode, whose frequency is given by $\omega\sim k_{\parallel}V_{A}$ when $k_{\perp}\rho_{s}\ll 1$,
lacks the parallel electric field to accelerate
the particles. A main objective of the present work is to emphasize that, on the contrary,
the role of the parallel electric field intrinsic to the Alfven wave
dynamics in a warm plasma should not be ignored, even at the scales of standard MHD.
We believe that while a scenario based on Alfven waves for electron
acceleration during solar flares remains a conjecture, as is any
other acceleration mechanism proposed so far, it is particulary attractive. The reason is due to the
 body of observational evidences on the role played by these waves in controlling many aspects of the
dynamics of astrophysical plasmas.

\begin{figure} \center
\includegraphics[width=89mm]{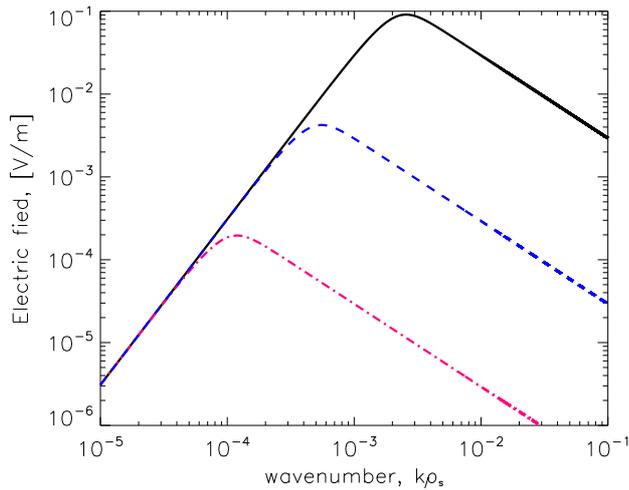}
\caption{Parallel electric field, $\delta E_{\parallel}$, as a function of $k_{\parallel}\rho_{s}$
for different values of the large scale magnetic field perturbation
$\epsilon=(\delta B_{\perp}/B_0)^2$, $1$- solid black line, $10^{-2}$- dash blue line,
$10^{-4}$ - dash-dot red line.}
\label{fig:Ep_k}
\end{figure}

In plasma conditions typical of solar flares, values of the fluctuating parallel electric field
can be significant. We consider a range of solar plasma parameters:
guiding magnetic field $B_0=100$~Gauss, plasma density $n_e=5\times 10^{9}$~cm$^{-3}$,
plasma temperature 1 MK, and the loop length scale $L_0$. Normalizing the wavenumber
by $\rho _s=C_{s}/\omega_{ci}$, the amplitude of the
parallel electric field is presented in Figure (\ref{fig:Ep_k}). For solar flare
parameters $\rho _s \simeq 10$ ~cm, the maximum electric fields $2\times 10^{-4}- 10^{-1}$ V$m^{-1}$
appears at the scales $\lambda = \rho _s/k_{\parallel} \sim 10^5 - 5\times 10^3$ cm.
The maximum values of the electric field and the characteristic scales can be derived
explicitly from the results of the previous section. The maximum of the electric field amplitude
\begin{equation}
\delta E_{\parallel}^{max}\sim \epsilon^{2/3}\rho^{2/3}
\label{eq:Emax}
\end{equation}
is reached at
\begin{equation}
k_{\parallel} ^{max}\sim\epsilon^{1/3}\rho_{s}^{-2/3}
\label{eq:kmax}
\end{equation}
These are dependent on the magnitude of the
magnetic perturbation $\epsilon=(\delta B_{\perp}/B_0)^2$ at the energy injection
scale $L_0$.

Our estimate for the maximum electric electric field fluctuation
produced by Alfven wave
turbulence show that the latter can be rather strong.
Indeed, for the adopted solar flares plasma parameters, the Dreicer electric field
is $\sim 0.02$ V$m^{-1}$, and the maximum amplitude of the fluctuating field
can exceed this values.

The parallel electric force associated with the Alfven wave dynamics could play an important role in a number
of cases. First, as the primary source
of thermal electron acceleration, where waves and turbulence are triggered
by the reconnection process\citep{Bellan1998,LongcopePriest2007} or as a result of the
twisting of the field lines anchored in the photosphere. It has also been proposed that part
of the energy released during magnetic reconnection is transported by Alfven waves to the chromosphere\citep{Emslie1982,Fletcher2008}.
Therefore, in a situation where Alfvenic turbulence fills the loop,
the waves will affect the transport of the energetic
electrons to the chromosphere. Acceleration occurs along the field lines which are perturbed by the
Alfven dynamics, hence the pace
of the acceleration along $\mathbf{B}=\mathbf{B}_{0}+\mathbf{\delta B}_{\perp}$ also controls the cross-$\mathbf{B}_{0}$
transport. Finally, as already mentioned in the Introduction, we note that if the electric field produced by Alfven turbulence
can re-accelerate non-thermal electrons injected into the chromosphere
this revision of the standard Thick Target Model may resolve existing problems with it. Whether this is really the case depends on
the detailed nature of the interaction between the parallel electric field and the electrons which requires a kinetic description,
 the subject of a future publication.

\begin{acknowledgements}

This work is supported by a STFC rolling grant (NHB, EPK, JCB) and an STFC Advanced
Fellowship (EPK). Financial support by the Leverhulme Trust
grant (F/00179/AY) and by the European Commission through the SOLAIRE
Network (MTRN-CT-2006-035484) is gratefully acknowledged.

\end{acknowledgements}

\bibliographystyle{aa}
\bibliography{refaa}

\begin{thebibliography}{52}
\expandafter\ifx\csname natexlab\endcsname\relax\def\natexlab#1{#1}\fi

\bibitem[{{Aschwanden}(2002)}]{Aschwanden2002}
{Aschwanden}, M.~J. 2002, Space Science Reviews, 101, 1

\bibitem[{{Aschwanden} {et~al.}(2002){Aschwanden}, {Brown}, \&
  {Kontar}}]{Aschwanden_etal2002}
{Aschwanden}, M.~J., {Brown}, J.~C., \& {Kontar}, E.~P. 2002, \solphys, 210,
  383

\bibitem[{{Bellan}(1998)}]{Bellan1998}
{Bellan}, P.~M. 1998, Physics of Plasmas, 5, 3081

\bibitem[{{Beresnyak} \& {Lazarian}(2006)}]{beres2006}
{Beresnyak}, A. \& {Lazarian}, A. 2006, \apjl, 640, L175

\bibitem[{{Beresnyak} \& {Lazarian}(2008)}]{Beresnyak2008}
{Beresnyak}, A. \& {Lazarian}, A. 2008, \apj, 682, 1070

\bibitem[{{Bian} \& {Tsiklauri}(2008)}]{Bian2008}
{Bian}, N. \& {Tsiklauri}, D. 2008, \aap, 489, 1291

\bibitem[{{Bian} \& {Tsiklauri}(2009)}]{BianTsiklauri2009}
{Bian}, N.~H. \& {Tsiklauri}, D. 2009, Physics of Plasmas, 16, 064503

\bibitem[{{Biskamp} {et~al.}(1999){Biskamp}, {Schwarz}, {Zeiler}, {Celani}, \&
  {Drake}}]{Biskamp_etal1999}
{Biskamp}, D., {Schwarz}, E., {Zeiler}, A., {Celani}, A., \& {Drake}, J.~F.
  1999, Physics of Plasmas, 6, 751

\bibitem[{{Bogachev} \& {Somov}(2007)}]{BogachevSomov2007}
{Bogachev}, S.~A. \& {Somov}, B.~V. 2007, Astronomy Letters, 33, 54

\bibitem[{{Boldyrev}(2006)}]{boldyrev2006}
{Boldyrev}, S. 2006, Physical Review Letters, 96, 115002

\bibitem[{{Brown}(1971)}]{Brown1971}
{Brown}, J.~C. 1971, \solphys, 18, 489

\bibitem[{{Brown}(1972)}]{Brown1972}
{Brown}, J.~C. 1972, \solphys, 26, 441

\bibitem[{{Brown} {et~al.}(2006){Brown}, {Emslie}, {Holman}, {Johns-Krull},
  {Kontar}, {Lin}, {Massone}, \& {Piana}}]{Brown_etal2006}
{Brown}, J.~C., {Emslie}, A.~G., {Holman}, G.~D., {et~al.} 2006, \apj, 643, 523

\bibitem[{{Brown} {et~al.}(2003){Brown}, {Emslie}, \&
  {Kontar}}]{Brown_etal2003}
{Brown}, J.~C., {Emslie}, A.~G., \& {Kontar}, E.~P. 2003, \apjl, 595, L115

\bibitem[{{Brown} \& {Kontar}(2005)}]{BrownKontar2005}
{Brown}, J.~C. \& {Kontar}, E.~P. 2005, Advances in Space Research, 35, 1675

\bibitem[{{Brown} {et~al.}(2009){Brown}, {Turkmani}, {Kontar}, {MacKinnon}, \&
  {Vlahos}}]{Brown_etal2009}
{Brown}, J.~C., {Turkmani}, R., {Kontar}, E.~P., {MacKinnon}, A.~L., \&
  {Vlahos}, L. 2009, \aap, 508, 993

\bibitem[{{Bykov} \& {Fleishman}(2009)}]{BykovFleishman2009}
{Bykov}, A.~M. \& {Fleishman}, G.~D. 2009, \apjl, 692, L45

\bibitem[{{Chandran}(2008)}]{Chandran2008}
{Chandran}, B.~D.~G. 2008, \apj, 685, 646

\bibitem[{{Cho} \& {Lazarian}(2004)}]{ChoLazarian2004}
{Cho}, J. \& {Lazarian}, A. 2004, \apjl, 615, L41

\bibitem[{{Cho} \& {Lazarian}(2009)}]{Cho2009}
{Cho}, J. \& {Lazarian}, A. 2009, \apj, 701, 236

\bibitem[{{Cranmer} \& {van Ballegooijen}(2003)}]{CranmerVanBallegooijen2003}
{Cranmer}, S.~R. \& {van Ballegooijen}, A.~A. 2003, \apj, 594, 573

\bibitem[{{Emslie} {et~al.}(2003){Emslie}, {Kontar}, {Krucker}, \&
  {Lin}}]{Emslie_etal2003}
{Emslie}, A.~G., {Kontar}, E.~P., {Krucker}, S., \& {Lin}, R.~P. 2003, \apjl,
  595, L107

\bibitem[{{Emslie} \& {Sturrock}(1982)}]{Emslie1982}
{Emslie}, A.~G. \& {Sturrock}, P.~A. 1982, \solphys, 80, 99

\bibitem[{{Fermi}(1949)}]{Fermi1949}
{Fermi}, E. 1949, Physical Review, 75, 1169

\bibitem[{{Fletcher} \& {Hudson}(2008)}]{Fletcher2008}
{Fletcher}, L. \& {Hudson}, H.~S. 2008, \apj, 675, 1645

\bibitem[{{Galtier} \& {Bhattacharjee}(2003)}]{Galtier2003}
{Galtier}, S. \& {Bhattacharjee}, A. 2003, Physics of Plasmas, 10, 3065

\bibitem[{{Galtier} {et~al.}(2002){Galtier}, {Nazarenko}, {Newell}, \&
  {Pouquet}}]{Galtier2002}
{Galtier}, S., {Nazarenko}, S.~V., {Newell}, A.~C., \& {Pouquet}, A. 2002,
  \apjl, 564, L49

\bibitem[{{Goldreich} \& {Sridhar}(1995)}]{GoldreichSridhar1995}
{Goldreich}, P. \& {Sridhar}, S. 1995, \apj, 438, 763

\bibitem[{{G{\'o}mez} {et~al.}(2008){G{\'o}mez}, {Mahajan}, \&
  {Dmitruk}}]{Gomez_etal2008}
{G{\'o}mez}, D.~O., {Mahajan}, S.~M., \& {Dmitruk}, P. 2008, Physics of
  Plasmas, 15, 102303

\bibitem[{{Holman}(1985)}]{Holman1985}
{Holman}, G.~D. 1985, \apj, 293, 584

\bibitem[{{Howes} {et~al.}(2008){Howes}, {Cowley}, {Dorland}, {Hammett},
  {Quataert}, \& {Schekochihin}}]{Howes_etal2008}
{Howes}, G.~G., {Cowley}, S.~C., {Dorland}, W., {et~al.} 2008, Journal of
  Geophysical Research (Space Physics), 113, 5103

\bibitem[{{Kadomtsev} \& {Pogutse}(1974)}]{KadomtsevPogutse1974}
{Kadomtsev}, B.~B. \& {Pogutse}, O.~P. 1974, Soviet Journal of Experimental and
  Theoretical Physics, 38, 283

\bibitem[{{Kontar} \& {Brown}(2006)}]{KontarBrown2006}
{Kontar}, E.~P. \& {Brown}, J.~C. 2006, \apjl, 653, L149

\bibitem[{{Kontar} {et~al.}(2008){Kontar}, {Hannah}, \&
  {MacKinnon}}]{Kontar_etal2008}
{Kontar}, E.~P., {Hannah}, I.~G., \& {MacKinnon}, A.~L. 2008, \aap, 489, L57

\bibitem[{{Kraichnan}(1965)}]{Kraichnan1965}
{Kraichnan}, R.~H. 1965, Physics of Fluids, 8, 1385

\bibitem[{{Krucker} \& {Lin}(2008)}]{KruckerLin2008}
{Krucker}, S. \& {Lin}, R.~P. 2008, \apj, 673, 1181

\bibitem[{{Lin} \& {Hudson}(1976)}]{LinHudson1976}
{Lin}, R.~P. \& {Hudson}, H.~S. 1976, \solphys, 50, 153

\bibitem[{{Lithwick} {et~al.}(2007){Lithwick}, {Goldreich}, \&
  {Sridhar}}]{Lithwick2007}
{Lithwick}, Y., {Goldreich}, P., \& {Sridhar}, S. 2007, \apj, 655, 269

\bibitem[{{Litvinenko}(2003)}]{Litv2003}
{Litvinenko}, Y.~E. 2003, Advances in Space Research, 32, 2385

\bibitem[{{Longcope} \& {Priest}(2007)}]{LongcopePriest2007}
{Longcope}, D.~W. \& {Priest}, E.~R. 2007, Physics of Plasmas, 14, 122905

\bibitem[{{Mason} {et~al.}(2006){Mason}, {Cattaneo}, \& {Boldyrev}}]{mason2006}
{Mason}, J., {Cattaneo}, F., \& {Boldyrev}, S. 2006, Physical Review Letters,
  97, 255002

\bibitem[{{Miller} {et~al.}(1997){Miller}, {Cargill}, {Emslie}, {Holman},
  {Dennis}, {LaRosa}, {Winglee}, {Benka}, \& {Tsuneta}}]{Miller_etal1997}
{Miller}, J.~A., {Cargill}, P.~J., {Emslie}, A.~G., {et~al.} 1997, \jgr, 102,
  14631

\bibitem[{{Miller} {et~al.}(1996){Miller}, {Larosa}, \&
  {Moore}}]{Miller_etal1996}
{Miller}, J.~A., {Larosa}, T.~N., \& {Moore}, R.~L. 1996, \apj, 461, 445

\bibitem[{{Ng} {et~al.}(2003){Ng}, {Bhattacharjee}, {Germaschewski}, \&
  {Galtier}}]{Ng_etal2003}
{Ng}, C.~S., {Bhattacharjee}, A., {Germaschewski}, K., \& {Galtier}, S. 2003,
  Physics of Plasmas, 10, 1954

\bibitem[{{Perez} \& {Boldyrev}(2009)}]{perez2009}
{Perez}, J.~C. \& {Boldyrev}, S. 2009, Physical Review Letters, 102, 025003

\bibitem[{{Petrosian}(1999)}]{Petrosian1999}
{Petrosian}, V. 1999, in Plasma Turbulence and Energetic Particles in
  Astrophysics, Proceedings of the International Conference, Cracow (Poland),
  5-10 September 1999, Eds.: Micha{\l} Ostrowski, Reinhard Schlickeiser,
  Obserwatorium Astronomiczne, Uniwersytet Jagiello{\'n}ski, Krak{\'o}w 1999,
  p. 135-146., ed. {M.~Ostrowski \& R.~Schlickeiser}, 135--146

\bibitem[{{Schekochihin} {et~al.}(2009){Schekochihin}, {Cowley}, {Dorland},
  {Hammett}, {Howes}, {Quataert}, \& {Tatsuno}}]{Schekochihin_etal2009}
{Schekochihin}, A.~A., {Cowley}, S.~C., {Dorland}, W., {et~al.} 2009, \apjs,
  182, 310

\bibitem[{{Siversky} \& {Zharkova}(2009)}]{SiverskyZharkova2009}
{Siversky}, T.~V. \& {Zharkova}, V.~V. 2009, Journal of Plasma Physics, 75, 619

\bibitem[{{Strauss}(1976)}]{Strauss1976}
{Strauss}, H.~R. 1976, Physics of Fluids, 19, 134

\bibitem[{{Syrovatskii} \& {Shmeleva}(1972)}]{Syrovatskii1972}
{Syrovatskii}, S.~I. \& {Shmeleva}, O.~P. 1972, \azh, 49, 334

\bibitem[{{Waelbroeck} {et~al.}(2009){Waelbroeck}, {Hazeltine}, \&
  {Morrison}}]{Waelbroeck_etal2009}
{Waelbroeck}, F.~L., {Hazeltine}, R.~D., \& {Morrison}, P.~J. 2009, Physics of
  Plasmas, 16, 032109

\bibitem[{{Wood} \& {Neukirch}(2005)}]{WoodNeukirch2005}
{Wood}, P. \& {Neukirch}, T. 2005, \solphys, 226, 73

\end{thebibliography}

\end{document}